\documentclass[aps,prb,twocolumn,nofootinbib,longbibliography,amsmath,amssymb]{revtex4-2}

\usepackage{graphicx}
\usepackage{bm}
\usepackage{dcolumn}
\usepackage[colorlinks=true,linkcolor=blue,citecolor=blue,urlcolor=blue]{hyperref}

\newcommand{\Pzz}{P_{zz}}
\newcommand{\wD}{\omega_D}
\newcommand{\wQ}{\omega_Q}
\newcommand{\wone}{\omega_1}

\newcommand{\dd}{\mathrm{d}}

\newcommand{\degree}{^\circ}

\begin{document}

\title{Coherent control of quadrupolar order in solid spin-1 systems:\
majorization bounds, orientation-indexed adiabatic inversion, and order lifetime}

\author{D. Keller}
\author{M. F. Hossain}
\affiliation{University of Virginia, Charlottesville, USA}


\begin{abstract}
Tensor polarization $\Pzz=1-3p_0$ of spin-1 nuclei in dynamically polarized solid targets is limited, under a single spin temperature, to $\Pzz^{\rm eq}=2-\sqrt{4-3P^2}$ ($\Pzz\simeq0.20$ at vector polarization $P=0.5$). Radio-frequency (rf) semi-saturation of one quadrupolar Pake-doublet branch, with and without slow sample rotation, has reached $\Pzz\simeq0.36$. Using the first-order quadrupole Hamiltonian, spin temperatures of the Zeeman and alignment (quadrupolar-order) reservoirs, and single-transition operators, we analyze these manipulations. Five results follow. (i) Under unital operations (rf pulses, rotation, saturation), $1-3\lambda_{\max}\le\Pzz\le1-3\lambda_{\min}$, where $\lambda_{\min,\max}$ are the smallest and largest initial populations. Saturation achieves at most half the available transfer; selective $\pi$ inversion reaches either bound ($\Pzz=-0.85,+0.65$ at $P=0.5$). (ii) In a powder, the two single-quantum transitions overlap except in the pedestal region. Adiabatic frequency sweeps confined there invert one transition per bond. Inverting each bond exactly once requires packing antipodal caps on the orientation sphere. The optimal single-axis arrangement has four stops at the magic angle $54.74\degree$ (cube body diagonals; 73\% coverage); six icosahedral axes give the global optimum (90\%). (iii) Rotation-induced Landau--Zener crossings of deuteron pairs are adiabatic at current rates and scramble the inversion pattern; transits must be fast for the strongest, geminal CD$_2$, pairs. (iv) A parity theorem determines the inverted bonds solely from the endpoints of continuous adiabatic passage: rf during rotation yields at most two caps, and a full turn none. (v) The enhanced state decays by dipolar flip-flops, with $T_{1Q}\le T_{1Z}/3$ and a duty-cycle break-even of $\simeq35$~min. Magic-angle spinning cannot replace this scheme.
\end{abstract}

\maketitle

\section{Introduction}

Spin-1 nuclei carry, beyond the vector polarization $P=\langle I_z\rangle$, an alignment or tensor polarization $\Pzz=\langle 3I_z^2-2\rangle$ that couples to the rank-2 structure of the nuclear ground state. In the deuteron this quantity controls the tensor structure function $b_1$ and the tensor analyzing powers of elastic and quasi-elastic scattering \cite{hoodbhoy1989,hermes2005,jlab_b1}. Solid dynamically polarized targets, deuterated alcohols and ammonia doped with paramagnetic centers and polarized at $1$~K and $5$~T \cite{crabb1997,goertz2002,abragamgoldman1978}, reach $P\simeq0.5$ for deuterons but, because dynamic nuclear polarization (DNP) establishes a single spin temperature across the nuclear Zeeman reservoir \cite{borghini1968,goldman1970}, deliver only the tensor polarization that spin temperature implies, $\Pzz\simeq0.2$.

Keller and co-workers have shown that this limit is not fundamental. By semi-saturating one branch of the quadrupolar doublet with an auxiliary rf field while DNP continues, and by slowly rotating the target to bring further bonds into reach of the rf, they obtained $\Pzz\simeq0.31$ without rotation and $\simeq0.36$ with it \cite{keller2017,keller2020,keller2026}, and the same group has since demonstrated adiabatic fast passage (AFP) of the full deuteron line in the same apparatus \cite{hossain2026}. These experiments pose a theoretical question that the present paper answers: what is the largest tensor polarization that rf manipulation and sample rotation can extract from a given DNP state, which sequence attains it, and what limits its lifetime?

The tools needed are classical. The quadrupole Hamiltonian and its powder spectrum follow Abragam \cite{abragam1961}, Cohen and Reif \cite{cohen1957} and Pake \cite{pake1948}; the thermodynamics of Zeeman and quadrupolar order follow Goldman \cite{goldman1970} and Jeener and Broekaert \cite{jeener1967}, with the deuteron-specific formulation of Spiess \cite{spiess1980}; selective manipulation of one transition of a multilevel spin uses the fictitious spin-$\tfrac12$ operators of Vega and Pines \cite{vega1977} and Wokaun and Ernst \cite{wokaun1977}; adiabatic passage follows Abragam \cite{abragam1961} and Slichter \cite{slichter1990} with the Landau--Zener formula \cite{landau1932,zener1932}; cross-relaxation and spectral spin diffusion follow Bloembergen \emph{et al.} \cite{bloembergen1959}, Provotorov \cite{provotorov1962} and Suter and Ernst \cite{suter1985}; and the analysis of sample spinning rests on the average Hamiltonian theory of Haeberlen and Waugh \cite{haeberlen1968} and Maricq and Waugh \cite{maricq1979}. The one ingredient that is not standard NMR is a packing problem on the sphere \cite{tammes1930,fejestoth1964,conway1996}, which arises because the selective inversion is itself a unitary operation and a unitary applied twice to the same bond undoes itself. The same fact, that an adiabatic passage traversed twice is the identity, governs what rotation can and cannot do when the rf is on, and the analysis of pair flip-flops in a rotating sample is the Landau--Zener problem applied to the dipolar coupling rather than to the rf, the same physics as the spin-locking adiabaticity of quadrupolar nuclei under sample spinning \cite{vega1992}.

The paper is organized as follows. Section~\ref{sec:system} sets up the spin-1 Hamiltonian, the chemistry of the deuteron electric field gradient (EFG) in the materials of interest, the powder lineshape, and the order parameters and their spin-temperature state. Section~\ref{sec:bound} proves the majorization bound. Section~\ref{sec:inversion} derives the selective adiabatic inversion and its pulse parameters. Section~\ref{sec:rotation} treats rotation as an orientation map and derives the stop angles. Section~\ref{sec:timing} derives the rotation-driven frequency modulation and pair crossings, their effect on an inversion pattern, the resulting timing window for the sequence, and a parity theorem for simultaneous rf and rotation. Section~\ref{sec:relaxation} derives the static decay channels of the enhanced state and the duty-cycle break-even. Section~\ref{sec:mas} shows why spinning and recoupling do not help. Section~\ref{sec:tests} proposes the measurements that decide the design, and Sec.~\ref{sec:conclusion} concludes.

\section{The spin-1 system in a solid}
\label{sec:system}

\subsection{Quadrupole Hamiltonian and the deuteron EFG}

A nucleus of spin $I=1$ with quadrupole moment $Q$ in a magnetic field $B_0\hat z$ and an electric field gradient (EFG) tensor $V_{\alpha\beta}=\partial^2V/\partial x_\alpha\partial x_\beta$ has the Hamiltonian \cite{abragam1961,cohen1957}
\begin{equation}
\begin{aligned}
H &= -\wD I_z + H_Q,\\
H_Q &= \frac{eQ}{6\hbar}\sum_{\alpha\beta}V_{\alpha\beta}\,\tfrac{3}{2}\bigl(I_\alpha I_\beta+I_\beta I_\alpha-\tfrac{2}{3}\delta_{\alpha\beta}I^2\bigr).
\end{aligned}
\end{equation}
with $\wD=\gamma_D B_0$ the Larmor frequency ($\wD/2\pi=32.7$~MHz at $5$~T). In the principal axis system of the EFG, with $eq=V_{zz}$ and asymmetry $\eta=(V_{xx}-V_{yy})/V_{zz}$,
\begin{equation}
H_Q=\frac{e^2qQ}{4\hbar}\Bigl[3I_{z'}^2-I(I+1)+\eta\bigl(I_{x'}^2-I_{y'}^2\bigr)\Bigr].
\end{equation}
At $5$~T the Zeeman term exceeds the quadrupole term by a factor of order $200$, and the secular (first-order) part of $H_Q$ in the laboratory frame is \cite{cohen1957}
\begin{equation}
\begin{aligned}
H_Q^{(1)}&=\wQ\,\bigl(3\cos^2\theta-1+\eta\sin^2\theta\cos2\phi\bigr)\,(3I_z^2-2),\\
\wQ&\equiv\frac{e^2qQ}{8\hbar}.
\end{aligned}
\label{eq:HQ1}
\end{equation}
where $(\theta,\phi)$ are the polar angles of $B_0$ in the EFG frame. The energies are $E_m=-\wD m+\wQ(3\cos^2\theta-1)(3m^2-2)$ for $\eta=0$, and the two allowed single-quantum transitions are
\begin{align}
\omega_{+1,0}&=\wD-3\wQ\,(3\cos^2\theta-1),\nonumber\\
\omega_{0,-1}&=\wD+3\wQ\,(3\cos^2\theta-1).
\label{eq:transitions}
\end{align}
The double-quantum transition $|{+1}\rangle\leftrightarrow|{-1}\rangle$ sits at $2\wD$ independent of $\wQ$ to first order. Second-order corrections are of order $\wQ^2/\wD\approx1$~kHz and are neglected except where noted.

For deuterium the EFG at the nucleus is dominated by the electron density of the covalent bond and the principal axis $z'$ lies, to within a few degrees, along the bond \cite{seelig1977,hoatson1994}. Aliphatic C--D bonds have $e^2qQ/h\approx165$--$175$~kHz with $\eta\lesssim0.05$; hydroxyl O--D bonds in hydrogen-bonded alcohols are larger, $\approx210$--$250$~kHz, with $\eta\approx0.1$; N--D bonds in ammonia are intermediate \cite{seelig1977,hoatson1994,schmidtrohr1994}. In the units of Eq.~\eqref{eq:HQ1} an aliphatic deuteron has $3\wQ/2\pi=\tfrac{3}{8}(e^2qQ/h)\approx64$~kHz, so the quadrupolar doublet spans $\pm2\times64$~kHz about $\wD$. Two features of the target chemistry matter for what follows.

First, the target materials are not single crystals. Deuterated butanol beads are quench-frozen glasses \cite{crabb1997,goertz2002}, so the distribution of bond directions is isotropic by construction and the EFG parameters carry a small glassy distribution; deuterated ammonia is polycrystalline with an isotropic distribution of crystallite orientations. In both cases the unit of the following analysis is the individual bond direction $\hat u$, not the crystallite: within one molecule the C--D bonds of a CD$_2$ group point in different directions and are treated separately.

Second, molecular motion is frozen at $1$~K except for quantum rotational tunnelling of methyl groups \cite{press1981,horsewill1999}. If a CD$_3$ group reorients about its $C_3$ axis fast compared with $\wQ$, its three deuterons share a motionally averaged EFG along the $C_3$ axis scaled by $P_2(\cos70.5\degree)=-\tfrac13$, giving $e^2qQ/h\approx55$~kHz and a second, narrower doublet. For CD$_3$ the tunnel splitting is several orders of magnitude smaller than for CH$_3$ at the same barrier \cite{horsewill1999} and, for the $\sim3$~kcal/mol barriers of alkyl chains, is well below $\wQ$, so the static limit applies; the sweep bands derived below must nevertheless be checked against the measured lineshape for a reduced-coupling component, since a site with a different $\wQ$ has its clean region at a different absolute frequency (Sec.~\ref{sec:tests}). In the experiments of Ref.~\cite{keller2020} the material was C$_4$D$_9$OH, so no O--D site is present and the single-site treatment is exact apart from the glassy distribution.

\subsection{Reduced frequency and the powder lineshape}

\begin{figure}
\includegraphics[width=\columnwidth]{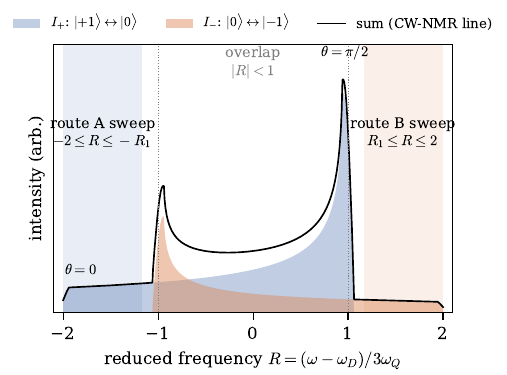}
\caption{Powder lineshape of a spin-1 nucleus with an axial EFG in the reduced frequency $R=(\omega-\wD)/3\wQ$, Eq.~\eqref{eq:pake}, with each branch weighted by its population difference at $P=0.5$ (Sec.~\ref{sec:order}). The horns at $R=\pm1$ correspond to $\theta=\pi/2$ and the pedestal edges at $R=\mp2$ to $\theta=0$. The two branches overlap for $|R|<1$; the shaded bands $R_1\le|R|\le2$ with $R_1=1.17$ are the clean regions swept by the adiabatic inversions of Sec.~\ref{sec:inversion} (route A on the $(+1,0)$ pedestal, route B on the $(0,-1)$ pedestal). A small Gaussian broadening has been applied to render the horns.}
\label{fig:line}
\end{figure}

It is convenient to measure frequency from $\wD$ in units of $3\wQ$,
\begin{equation}
\begin{aligned}
R&\equiv\frac{\omega-\wD}{3\wQ},\\
R_{+1,0}&=-(3\cos^2\theta-1),\quad R_{0,-1}=+(3\cos^2\theta-1).
\end{aligned}
\label{eq:R}
\end{equation}
Since $3\cos^2\theta-1\in[-1,2]$, the $(+1,0)$ branch occupies $R\in[-2,1]$ and the $(0,-1)$ branch $R\in[-1,2]$. For an isotropic distribution of bond directions the number of bonds per unit $R$ on each branch is, with $\dd(\cos\theta)$ uniform,
\begin{equation}
g_\pm(R)=\frac{1}{2\sqrt{3(1\mp R)}},
\qquad \int g_\pm\,\dd R=1,
\label{eq:pake}
\end{equation}
the Pake density \cite{pake1948}: singular horns at $R=\pm1$ ($\theta=\pi/2$, the most probable orientation) and flat pedestals ending at $R=\mp2$ ($\theta=0$). Figure~\ref{fig:line} shows the two branches weighted by their population differences. Three consequences of Eq.~\eqref{eq:R} carry the rest of the paper:
\begin{enumerate}
\item The branches overlap for $|R|<1$. At any such frequency both transitions are present, from different bonds.
\item For $R>1$ only the $(0,-1)$ branch exists and for $R<-1$ only the $(+1,0)$ branch. A bond contributes to these \emph{clean regions} if and only if
\begin{equation}
\cos^2\theta>\tfrac23,\quad\text{i.e.}\quad \theta<\theta_c=\arccos\sqrt{\tfrac23}=35.26\degree.
\label{eq:thetac}
\end{equation}
\item The two transitions of one bond are mirror images about $\wD$, separated by $2|R|\,3\wQ\ge6\wQ$ for a bond in the clean region, $\approx130$~kHz for C--D. Driving one of them does not perturb the other.
\end{enumerate}

\subsection{Order parameters and the spin-temperature state}
\label{sec:order}

For a spin-1 ensemble with populations $p_+,p_0,p_-$ of the Zeeman levels ($\sum p_m=1$) the vector and tensor polarizations are
\begin{equation}
\begin{aligned}
P&=\langle I_z\rangle=p_+-p_-,\\
\Pzz&=\langle 3I_z^2-2\rangle=p_+-2p_0+p_-=1-3p_0.
\end{aligned}
\label{eq:PPzz}
\end{equation}
with ranges $-1\le P\le1$ and $-2\le\Pzz\le1$. In the language of spin thermodynamics \cite{goldman1970,abragamgoldman1982}, $\langle I_z\rangle$ is the Zeeman order and $\langle 3I_z^2-2\rangle$ the alignment or quadrupolar order; Jeener and Broekaert showed how a pair of pulses converts one into the other in dipolar solids \cite{jeener1967}, and Spiess developed the deuteron version, spin alignment, as a probe of slow motion \cite{spiess1980}. Any diagonal density matrix of a spin-1 is a two-temperature state,
\begin{equation}
p_m\propto\exp\bigl[\,a\,m+b\,(3m^2-2)\,\bigr],
\label{eq:twotemp}
\end{equation}
since three populations are fixed by $a$, $b$ and normalization; $a$ is an inverse Zeeman temperature and $b$ an inverse alignment temperature. The single-temperature manifold $b=0$ is characterized by
\begin{equation}
p_0^2=p_+p_-,
\label{eq:detbal}
\end{equation}
and on it, writing $r=e^{a}$, $P=(r^2-1)/(r^2+r+1)$ and $\Pzz=(r^2-2r+1)/(r^2+r+1)$ \cite{dulya1997}; eliminating $r$,
\begin{equation}
\Pzz^{\rm eq}(P)=2-\sqrt{4-3P^2}.
\label{eq:Pzzeq}
\end{equation}

Thermal equilibrium with the lattice is a single-temperature state because $\hbar\wQ/k_BT\sim10^{-6}$ at $1$~K, and the DNP steady state is one as well: in the thermal-mixing picture \cite{borghini1968,abragamgoldman1978,wenckebach2016} the nuclear Zeeman reservoir is cooled by contact with the electron dipolar reservoir, while nothing cools the alignment reservoir, whose energy scale $\hbar\wQ$ is negligible. The measured deuteron polarizations $P\simeq0.5$ at $1$~K and $5$~T \cite{keller2020} therefore come with $\Pzz\simeq0.197$ and populations
\begin{equation}
(p_+,p_0,p_-)=(0.616,\,0.268,\,0.116),
\label{eq:pops}
\end{equation}
which are the input to everything that follows. Equation~\eqref{eq:detbal} will reappear in Sec.~\ref{sec:relaxation} as the fixed point of the dominant decay channel: the enhanced state relaxes back onto the curve \eqref{eq:Pzzeq}, not to zero.

The NMR observable is the absorption line, whose two branches have intensities $I_+(R)\propto(p_+-p_0)\,g_+(R)$ and $I_-(R)\propto(p_0-p_-)\,g_-(R)$ for a bond at the corresponding $\theta$; integrated over the powder, $P\propto I_++I_-$ and $\Pzz\propto I_+-I_-$, relations that hold for any population distribution \cite{dulya1997,keller2013}, whereas the intensity-ratio formulas hold only on the single-temperature manifold.

\section{A bound on coherent and saturating manipulations}
\label{sec:bound}

Every operation available to an NMR experimentalist short of contact with a bath, rf pulses of any shape, free evolution, and mechanical reorientation of the sample, which changes $\theta$ in Eq.~\eqref{eq:HQ1} but leaves the evolution unitary, acts on the density matrix of a bond as $\rho\mapsto U\rho U^\dagger$. Saturation, the incoherent limit of rf irradiation, acts as a convex combination of such unitaries. All of these are unital maps, $\Phi(\mathbf{1})=\mathbf{1}$. The following statement bounds all of them at once.

\paragraph*{Proposition.} Let $\rho$ have eigenvalues $\lambda_1\ge\lambda_2\ge\lambda_3$ and let $\Phi$ be unital. Then the Zeeman-basis populations $p=\mathrm{diag}\,\Phi(\rho)$ satisfy $p\prec\lambda$ (majorization), and in particular $\lambda_3\le p_0\le\lambda_1$, so that
\begin{equation}
1-3\lambda_{\max}\;\le\;\Pzz\;\le\;1-3\lambda_{\min}.
\label{eq:bound}
\end{equation}
Both ends are attained, and only by permutations of the eigenvalues into the Zeeman basis.

\paragraph*{Proof.} For $\Phi$ unitary, $\mathrm{diag}(U\rho U^\dagger)\prec\lambda(\rho)$ is the Schur--Horn theorem \cite{schur1923,horn1954,marshall2011}. For a convex combination of unitaries the diagonal is the same convex combination of vectors each majorized by $\lambda$, and the set $\{p:p\prec\lambda\}$ is convex; the general unital case follows from the Uhlmann theorem that the spectrum of $\Phi(\rho)$ is majorized by that of $\rho$ for unital $\Phi$ \cite{marshall2011}. Majorization implies $\lambda_{\min}\le p_i\le\lambda_{\max}$ for each component, which with $\Pzz=1-3p_0$ gives Eq.~\eqref{eq:bound}. Equality in $p_0=\lambda_{\min}$ requires $p$ to be a vertex of the permutohedron of $\lambda$, i.e.\ a permutation, which is reached by a unitary that maps eigenvectors to Zeeman states. $\square$

If $\rho$ is diagonal to begin with, as the DNP state is, then $\lambda=(p_+,p_0,p_-)$ and the permutations that attain the bound are transpositions of adjacent levels, each a selective $\pi$ rotation of one transition. Table~\ref{tab:bound} lists the possibilities from the state \eqref{eq:pops}. The asymmetry of the bound is the asymmetry of the range of $\Pzz$: emptying $m=0$ can raise $\Pzz$ by at most $1-0.197=0.80$, filling it can lower $\Pzz$ by up to $2.20$. Route A, which fills $m=0$ with the largest population, is the larger effect and leaves a small residual vector polarization, which is itself useful for separating tensor from vector observables.

\begin{table}
\caption{Population manipulations of one transition from the DNP state \eqref{eq:pops}. Saturation equalizes the two levels; a $\pi$ rotation exchanges them. The saturation values explain the ceiling of the semi-saturation experiments of Ref.~\cite{keller2020} ($\Pzz=0.36$ measured against a ceiling of $0.42$); the $\pi$-rotation values are the bound \eqref{eq:bound}.}
\label{tab:bound}
\begin{ruledtabular}
\begin{tabular}{lccc}
Operation & $(p_+,p_0,p_-)$ & $\Pzz$ & $P$ \\
\hline
none (DNP) & $(.616,.268,.116)$ & $+0.197$ & $0.500$ \\
sat.\ $(+1,0)$ & $(.442,.442,.116)$ & $-0.326$ & $0.326$ \\
sat.\ $(0,-1)$ & $(.616,.192,.192)$ & $+0.424$ & $0.424$ \\
A: $\pi$ on $(+1,0)$ & $(.268,.616,.116)$ & $-0.849$ & $0.151$ \\
B: $\pi$ on $(0,-1)$ & $(.616,.116,.268)$ & $+0.652$ & $0.349$ \\
\end{tabular}
\end{ruledtabular}
\end{table}

Saturation is the $t=\tfrac12$ member of the family of T-transforms that generate majorization \cite{marshall2011} and transfers exactly half of the population that the corresponding transposition does; this is why no refinement of the rf power profile of a semi-saturation scheme can pass the saturation ceiling, and why the measured $0.36$ of Ref.~\cite{keller2020} sits just below the $0.42$ of Table~\ref{tab:bound}. Exceeding the bound \eqref{eq:bound} requires a non-unital step, contact with a bath that changes the spectrum of $\rho$. In the target that bath is DNP itself: alternating DNP contact with selective transpositions is the heat-bath algorithmic cooling of Boykin \emph{et al.} \cite{boykin2002,fernandez2004} applied to a three-level system, and its asymptotic value is set by the electron polarization rather than by $\lambda_{\min}$. The remainder of this paper treats a single cycle; the cycle itself is discussed in Sec.~\ref{sec:relaxation}.

\section{Selective adiabatic inversion of one branch}
\label{sec:inversion}

\subsection{Single-transition operators}

In the basis $\{|{+1}\rangle,|0\rangle,|{-1}\rangle\}$ the rf Hamiltonian in the frame rotating at $\omega_{\rm rf}$ is $\wone I_x$ with $\wone=\gamma_D B_1$ the rotating-field amplitude, and
\begin{equation}
I_x=\sqrt2\,\bigl(I_x^{(12)}+I_x^{(23)}\bigr),
\end{equation}
where $I_\alpha^{(rs)}$ are the fictitious spin-$\tfrac12$ operators of the transition $r\leftrightarrow s$ \cite{vega1977,wokaun1977}. Projected onto the $(0,-1)$ pair (route B; route A on the $(+1,0)$ pair is identical) the rotating-frame Hamiltonian of a bond at angle $\theta$ is
\begin{equation}
H_{\rm eff}(t)=\delta(t)\,I_z^{(23)}+\sqrt2\,\wone\,I_x^{(23)},\qquad
\delta(t)=\omega_{\rm rf}(t)-\omega_{0,-1}(\theta),
\label{eq:Heff}
\end{equation}
an isolated two-level system with effective Rabi frequency $\Omega_1=\sqrt2\,\wone$. The neglected $(12)$ term is detuned by $2|R|\,3\wQ\ge6\wQ$ and contributes only a Bloch--Siegert shift \cite{bloch1940} of order $\wone^2/6\wQ$, below $10$~Hz for the parameters below; second-order double-quantum nutation \cite{vega1977} scales as $\wone^2/3\wQ$ and is likewise negligible. The isolation of one transition is thus exact to the order required, provided the rf never enters the overlap region $|R|<1$, where the same frequency would address both transitions of different bonds.

\subsection{Adiabatic passage and the Landau--Zener condition}

A frequency sweep through resonance inverts the fictitious spin if the effective field rotates slowly compared with the precession about it \cite{abragam1961,slichter1990}, $|\dd\vartheta/\dd t|\ll\sqrt{\delta^2+\Omega_1^2}$ with $\tan\vartheta=\Omega_1/\delta$. For a linear sweep $\dd\omega_{\rm rf}/\dd t=\alpha$ the non-adiabatic (diabatic) probability is the Landau--Zener result \cite{landau1932,zener1932}
\begin{equation}
P_{\rm dia}=\exp\Bigl(-\frac{2\pi(\Omega_1/2)^2}{\alpha}\Bigr)=\exp\Bigl(-\pi\frac{\wone^2}{\alpha}\Bigr)\equiv e^{-\pi\mathcal{A}},
\label{eq:LZ}
\end{equation}
with adiabaticity parameter $\mathcal{A}=\wone^2/\alpha$. Inversion fidelity is $95.7\%$ at $\mathcal{A}=1$ and $99.8\%$ at $\mathcal{A}=2$, independent of sweep direction and of the precise value of $\wone$ once $\mathcal{A}\gtrsim2$. This insensitivity is the reason to use a swept pulse rather than a hard $\pi$ pulse: the rf field of a coil wound around a $15$~cm$^3$ target cell varies by tens of percent over the sample, a hard pulse would then fail at the level of the variation, while an adiabatic pulse tolerates it \cite{baum1985,garwood2001}. Amplitude- and frequency-modulated shapes, the hyperbolic secant of Silver \emph{et al.} \cite{silver1984}, the WURST family \cite{kupce1995} and their relatives \cite{tannus1997}, confine the inversion to a sharply bounded band, which is what the clean-region condition demands.

\subsection{Pulse specification}

Let the sweep cover the band $R_1\le|R|\le2$ with $R_1\ge1$; the bandwidth is $\Delta\omega=(2-R_1)\,3\wQ$. For a linear chirp $\alpha=\Delta\omega/T_p$ and the pulse length that gives adiabaticity $\mathcal{A}$ is
\begin{equation}
T_p=\frac{\mathcal{A}\,\Delta\omega}{\wone^2}=\frac{\mathcal{A}\,\mathrm{BW}}{2\pi\,\nu_1^2},
\label{eq:Tp}
\end{equation}
with BW and $\nu_1=\wone/2\pi$ in Hz. A hyperbolic-secant pulse, $\wone(t)=\wone^{\max}{\rm sech}(\beta\tau)$, $\delta(t)=\tfrac12\Delta\omega\tanh(\beta\tau)$, $\tau=2t/T_p-1$, $\beta\simeq5.3$ \cite{silver1984}, sweeps fastest at its center and needs $T_p=\mathcal{A}\beta\Delta\omega/(\wone^{\max})^2$ for the same central adiabaticity, about five times longer, in exchange for sharper band edges. Three further conditions bracket the pulse:
\begin{enumerate}
\item $T_p\ll T_{1}$ and $T_{1\rho}$, so that no relaxation occurs during the sweep; trivially satisfied at $1$~K.
\item The time a bond spends inside its own resonance, $\tau_p\simeq2\Omega_1/\alpha=2\sqrt2\,\mathcal{A}/\wone$, should not exceed the transverse dephasing time set by the deuteron--deuteron dipolar coupling. For nearest neighbours at $r\simeq1.8$~\AA{} the coupling is $b=(\mu_0/4\pi)\gamma_D^2\hbar/r^3\simeq2\pi\times0.5$~kHz, so $T_2$ is of order $0.3$--$1$~ms; the illustrative values $T_2=50$--$100~\mu$s quoted in Ref.~\cite{keller2026} for ND$_3$ would call for a larger $\wone$. During the passage the magnetization is spin-locked along the effective field and the relevant time is $T_{1\rho}\ge T_2$ \cite{slichter1990}.
\item The band edge. Bonds whose transition lies within $\simeq\Omega_1/3\wQ$ of $R_1$ in reduced units are partially inverted. With the density \eqref{eq:pake} this is a fraction $\approx g(R_1)\,\Omega_1/3\wQ\approx1.5\%$ of the spins per band edge for $\nu_1=5$~kHz; Sec.~\ref{sec:rotation} places these bonds where neighbouring caps touch, so the loss is counted once.
\end{enumerate}

Table~\ref{tab:pulse} collects the derived parameters for the apparatus of Refs.~\cite{keller2020,hossain2026}: a $20$-turn coil of diameter $2.5$~cm and length $3$~cm ($L\approx8~\mu$H) driven by a $50$~W amplifier, for which $B_1\simeq\tfrac12\mu_0NI/\ell$ with $I=\sqrt{2PQ_{\rm coil}/\omega L}$ gives $\nu_1\approx2$--$5$~kHz for $Q_{\rm coil}=20$--$50$. Reference~\cite{hossain2026} reports full-line deuteron AFP in this system with a $400$~kHz sweep in $0.2$~s, $\alpha=1.26\times10^7$~rad\,s$^{-2}$, with reversal efficiency $0.80\pm0.03$ in irradiated d-butanol; a sweep of only the $70$~kHz clean band at the same $\alpha$ takes $35$~ms and should do better, since it never crosses the overlap region where a full-line sweep addresses both transitions of the same bonds in succession.

\begin{table*}
\caption{Pulse parameters for the clean-band sweep, with $3\wQ/2\pi=64$~kHz (C--D, $e^2qQ/h=170$~kHz) and the coil of Ref.~\cite{keller2020}.}
\label{tab:pulse}
\begin{ruledtabular}
\begin{tabular}{lll}
Quantity & Value & Basis \\
\hline
band, route A & $-2\le R\le-1.17$ & $(+1,0)$ pedestal \\
band, route B & $+1.17\le R\le2$ & $(0,-1)$ pedestal \\
bandwidth & $53$~kHz ($70$~kHz at $3\wQ/2\pi=85$~kHz) & $(2-R_1)\,3\wQ$ \\
$\nu_1$ & $2$--$5$~kHz & $20$--$50$~W, $Q_{\rm coil}$ $20$--$50$ \\
$T_p$, linear, $\mathcal{A}=2$ & $0.7$--$2$~ms & Eq.~\eqref{eq:Tp} \\
$T_p$, sech, $\mathcal{A}=2$ & $4$--$10$~ms & $\mathcal{A}\beta\,\Delta\omega/\wone^2$ \\
$\tau_p$ per bond & $0.2$~ms at $\nu_1=5$~kHz & $2\sqrt2\mathcal{A}/\wone$ \\
fidelity & $>99.8\%$ & $1-e^{-2\pi}$ \\
energy per pulse & $0.05$--$0.5$~J & $50$~W$\times T_p$ \\
\end{tabular}
\end{ruledtabular}
\end{table*}

\section{Sample rotation as an orientation map}
\label{sec:rotation}

\subsection{The map and the cap}

Let $\hat u$ be a bond direction and $\hat n$ the direction of $B_0$, both in the sample frame, so that $\cos\theta=\hat u\cdot\hat n$. Rotating the sample by $\varphi$ about a laboratory axis $\hat a$ at angle $\beta$ to $B_0$ is equivalent to moving $\hat n$ on a cone of half-angle $\beta$ about $\hat a$,
\begin{equation}
\hat n(\varphi)=\cos\beta\,\hat a+\sin\beta\,(\cos\varphi\,\hat e_1+\sin\varphi\,\hat e_2),
\label{eq:cone}
\end{equation}
with $(\hat e_1,\hat e_2,\hat a)$ orthonormal. A single rotation axis places $B_0$ only on that cone; any other direction requires a second degree of freedom.

By Eq.~\eqref{eq:thetac}, a sweep over $R_1\le|R|\le2$ inverts one transition of every bond with $|\hat u\cdot\hat n|\ge\cos\theta_1$,
\begin{equation}
\cos^2\theta_1=\frac{1+R_1}{3},\qquad \theta_1\le\theta_c,
\label{eq:theta1}
\end{equation}
that is, a pair of antipodal caps of half-angle $\theta_1$ about $\pm\hat n$ with fractional area $1-\cos\theta_1$ of the unit sphere. Because only $|\cos\theta|$ enters, a stop is a line through the origin, not a vector.

\subsection{The hit-once condition and cap packing}

A $\pi$ rotation applied twice to the same bond restores its populations. For $N$ stop lines $\hat n_1,\dots,\hat n_N$ the caps must therefore be pairwise disjoint: two lines at angle $s$ (the smaller of the angle and its supplement) have disjoint caps when $s\ge2\theta_1$. The fraction of bonds inverted exactly once is then
\begin{equation}
\begin{aligned}
F(N,\theta_1)&=N\,(1-\cos\theta_1),\\
2\theta_1&\le s_{\min}(\hat n_1,\dots,\hat n_N),\quad\theta_1\le\theta_c .
\end{aligned}
\label{eq:F}
\end{equation}
Maximizing $F$ is the problem of packing $N$ lines in $\mathbb{R}^3$ with the largest minimum angle, the real projective version of the Tammes problem \cite{tammes1930,fejestoth1964,conway1996}, with the additional ceiling $\theta_c$ imposed by the lineshape. Its small-$N$ solutions are classical and give the stop angles directly.

\paragraph*{Single axis, four stops.} For $N$ stops equally spaced on the cone \eqref{eq:cone} adjacent stops are separated by
\begin{equation}
\cos s_{\rm adj}=\cos^2\beta+\sin^2\beta\cos(2\pi/N).
\label{eq:sadj}
\end{equation}
With $N=4$ the azimuthal cosine vanishes and $s_{\rm adj}=\arccos(\cos^2\beta)$. Using the full cap $\theta_1=\theta_c$ requires $s_{\rm adj}=2\theta_c$, and $\cos2\theta_c=2\cos^2\theta_c-1=\tfrac13$, so
\begin{equation}
\cos^2\beta=\tfrac13,\qquad\beta=54.74\degree .
\end{equation}
The magic angle appears here from the lineshape alone: the clean-region condition fixes the cap, the cap fixes the separation, and only the magic-angle cone holds four lines at that separation. The four stops are the body diagonals of a cube, mutually at $70.53\degree$; second neighbours are at $109.47\degree$, whose supplement is again $70.53\degree$, so no antipodal collision occurs. The coverage is $F=4(1-\sqrt{2/3})=0.734$.

\paragraph*{Single axis, five stops.} Second neighbours now constrain the packing through their antipodes: both $s_{\rm adj}$ and $180\degree-s_{\rm 2nd}$ must exceed $2\theta_1$. They are equal at $\cos^2\beta=\tfrac15$, $\beta=63.43\degree$, where $s_{\rm adj}=63.43\degree$ and $s_{\rm 2nd}=116.57\degree$. The stops are five of the six axes of an icosahedron, $\theta_1=31.72\degree$, $R_1=3\cos^2\theta_1-1=1.171$ and $F=5(1-\cos31.72\degree)=0.747$. The band is the outer $83\%$ of the pedestal, which also keeps the sweep edge further from the overlap region.

\begin{figure}
\includegraphics[width=\columnwidth]{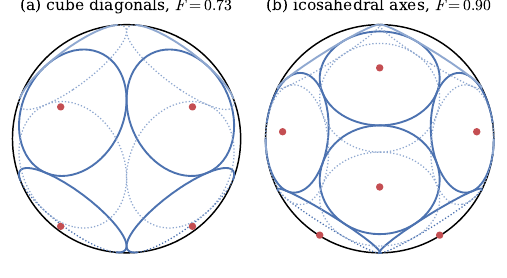}
\caption{Orthographic view of the orientation sphere with the antipodal caps of half-angle $\theta_1$ inverted at each stop (solid, front hemisphere; dotted, back). (a) Four stops on a cone at the magic angle $\beta=54.74\degree$, $\theta_1=\theta_c=35.26\degree$: the cube body diagonals, $F=0.73$. (b) Five stops on a cone at $\beta=63.43\degree$ plus the pole, $\theta_1=31.72\degree$: the icosahedral axes, $F=0.90$. In both the caps are tangent and no bond is inverted twice.}
\label{fig:caps}
\end{figure}

\paragraph*{Two axes, six stops.} The pole $\hat a$ is at $\beta=63.43\degree=2\theta_1$ from every cone stop, so it can be added as a sixth stop at no cost in separation. The six lines are the icosahedral axes, the optimal packing of six lines in $\mathbb{R}^3$ \cite{fejestoth1964,conway1996}, and $F=6(1-\cos31.72\degree)=0.896$. Seven lines pack at $s_{\min}\simeq53\degree$, giving $\theta_1=26.6\degree$ and $F=0.74$, and the hexagonal limit for many small caps is $\pi/2\sqrt3=0.907$ \cite{fejestoth1964}; the six-stop scheme therefore captures essentially the whole geometric ceiling. Figure~\ref{fig:caps} shows the two single- and two-axis solutions, and Table~\ref{tab:caps} lists the coverage, which was also checked by Monte Carlo integration over $3\times10^5$ isotropic bond directions (Appendix~\ref{app:mc}).

\begin{table*}
\caption{Stop geometries and coverage. $s_{\min}$ is the minimum angle between stop lines (antipodes included). The last two rows are the geometry of Ref.~\cite{keller2020}, a rotation axis perpendicular to $B_0$.}
\label{tab:caps}
\begin{ruledtabular}
\begin{tabular}{lccccc}
Scheme & $\beta$ & stops & $\theta_1$ & $R_1$ & $F$ \\
\hline
cube diagonals & $54.74\degree$ & $4\times90\degree$ & $35.26\degree$ & $1.00$ & $0.734$ \\
five icosahedral & $63.43\degree$ & $5\times72\degree$ & $31.72\degree$ & $1.17$ & $0.747$ \\
six icosahedral & $63.43\degree$+pole & $5\times72\degree+1$ & $31.72\degree$ & $1.17$ & $0.896$ \\
perpendicular, stepped & $90\degree$ & $4\times45\degree$ & $35.26\degree$ & $1.00$ & $0$ (37\% hit twice) \\
perpendicular, continuous & $90\degree$ & -- & -- & -- & $0.58$ reachable \\
\end{tabular}
\end{ruledtabular}
\end{table*}

\subsection{Why continuous rotation about a perpendicular axis cannot do this}

With $\beta=90\degree$ all stops lie on a great circle. Four stops at $45\degree$ have $s_{\min}=45\degree<2\theta_c$ and the Monte Carlo gives no bond inverted exactly once and $37\%$ inverted twice. Under continuous rotation about that axis, Eq.~\eqref{eq:cone} gives $\cos\theta(\varphi)=\sin\alpha_u\cos(\varphi+\varphi_0)$ for a bond at angle $\alpha_u$ to the axis, so a bond ever reaches the clean region only if $\sin\alpha_u\ge\cos\theta_c$, i.e.\ $\alpha_u\ge54.74\degree$: $58\%$ of the bonds, with $42\%$ permanently out of reach. This is a limit of the geometry used in Ref.~\cite{keller2020} that no rf refinement removes; a magic-angle axis raises the reachable fraction to $94\%$ under continuous rotation, but every reachable bond then traverses the band twice per turn, which is why the stops must be discrete and why rotation in this scheme is a positioning step rather than a transport mechanism (Sec.~\ref{sec:parity} makes this precise).

\section{Rotation-driven frequency modulation, spin diffusion, and the timing of the sequence}
\label{sec:timing}

Rotation does two things besides positioning the sample. It modulates every transition frequency in the powder, and in doing so it sweeps pairs of deuterons through mutual resonances that a static sample never reaches, turning dipolar flip-flops that are energetically forbidden at rest into adiabatic level crossings. This section derives both effects from the pair Hamiltonian, quantifies what they do to an inversion pattern, and shows that they impose a definite timing structure on the sequence: the sample must be static during each sweep and must move fast between sweeps. It then proves that applying the rf during the motion cannot improve on that structure.

\subsection{Frequency modulation of a single bond}
\label{sec:modulation}

Let the bond $\hat u$ make polar angle $\alpha_u$ with the rotation axis $\hat a$ and let $\phi_u$ be its azimuth in the sample frame. From Eq.~\eqref{eq:cone}, with $\varphi=\Omega t$,
\begin{equation}
\cos\theta(t)=\cos\beta\cos\alpha_u+\sin\beta\sin\alpha_u\cos(\Omega t+\phi_u),
\label{eq:costheta_t}
\end{equation}
and the reduced quadrupole coordinate $c(t)\equiv3\cos^2\theta(t)-1$, in terms of which $\omega_{+1,0}=\wD-3\wQ c$ and $\omega_{0,-1}=\wD+3\wQ c$, evolves as
\begin{equation}
\dot c=-6\,\Omega\,\sin\beta\sin\alpha_u\cos\theta\,\sin(\Omega t+\phi_u),\qquad |\dot c|\le6\,\Omega\sin\beta .
\label{eq:cdot}
\end{equation}
Each transition frequency is therefore chirped by the rotation at a rate bounded by
\begin{equation}
\alpha_{\rm rot}\equiv3\wQ\,\max|\dot c|=18\,\wQ\,\Omega\sin\beta ,
\label{eq:alpharot}
\end{equation}
which for C--D ($\wQ/2\pi=21$~kHz) and $\beta=54.74\degree$ is $\alpha_{\rm rot}=1.96\times10^{6}\,\Omega$~rad\,s$^{-2}$ with $\Omega$ in rad\,s$^{-1}$. At the $40$~s period of Ref.~\cite{keller2020} this is $3\times10^{5}$~rad\,s$^{-2}$, about $50$~kHz\,s$^{-1}$; a bond traverses the full width of its branch, $3\times3\wQ$, in a fraction of a turn. Over one turn $c(t)$ is periodic and every bond with $\alpha_u$ in the band $|\beta-\theta_1|<\alpha_u<\beta+\theta_1$ enters and leaves the cap of Sec.~\ref{sec:rotation} once; this is the geometric fact behind both the double-traversal remark of Sec.~\ref{sec:rotation} and the parity theorem of Sec.~\ref{sec:parity}.

\subsection{Pair flip-flops as level crossings}
\label{sec:paircrossing}

The secular dipolar coupling between deuterons $i$ and $j$ \cite{abragam1961},
\begin{align}
H_D^{ij}&=b_{ij}\Bigl[I_{zi}I_{zj}-\tfrac14\bigl(I_i^+I_j^-+I_i^-I_j^+\bigr)\Bigr],\nonumber\\
b_{ij}&=\frac{\mu_0}{4\pi}\frac{\gamma_D^2\hbar}{r_{ij}^3}\bigl(1-3\cos^2\vartheta_{ij}\bigr),
\label{eq:HD}
\end{align}
conserves $I_{zi}+I_{zj}$ and connects product states that differ by one unit of $m$ transferred between the spins. The quadrupole energies of Eq.~\eqref{eq:transitions} split such pairs of states by amounts that depend on the two bond orientations, and the flip-flop is effective only when that splitting is comparable to the coupling. For spin-1 pairs there are two classes of connected states (Appendix~\ref{app:pair}):
\begin{align}
\text{same branch:}\quad&|{+1},0\rangle\leftrightarrow|0,{+1}\rangle,\qquad \Delta^{\rm s}_{ij}=-3\wQ\,(c_i-c_j),\nonumber\\
\text{cross branch:}\quad&|0,0\rangle\leftrightarrow|{+1},{-1}\rangle,\ |{-1},{+1}\rangle,\qquad \Delta^{\rm c}_{ij}=3\wQ\,(c_i+c_j),
\label{eq:Deltas}
\end{align}
and the analogous $|0,-1\rangle\leftrightarrow|-1,0\rangle$ pair on the lower branch. In each two-dimensional subspace the matrix element of the flip-flop term is $-\tfrac14b_{ij}\times2=-\tfrac12b_{ij}$, so the pair is a two-level system
\begin{equation}
H_{\rm pair}(t)=\tfrac12\Delta_{ij}(t)\,\sigma_z+V_{ij}\,\sigma_x,\qquad V_{ij}=\tfrac12|b_{ij}| ,
\label{eq:Hpair}
\end{equation}
with $V_{ij}/2\pi\simeq240$~Hz for a geminal CD$_2$ pair ($r=1.8$~\AA, angular factor set to unity), $90$~Hz at $2.5$~\AA{} and $50$~Hz at $3$~\AA.

\paragraph*{Static sample.} $\Delta_{ij}$ is fixed and the flip-flop proceeds only if $|\Delta_{ij}|\lesssim$ the local dipolar width, a condition met for a fraction of order $V/3\wQ\sim10^{-2}$ of all pairs. Nearest neighbours on one molecule have fixed relative orientation, so for most orientations of the molecule none of its internal pairs is matched. This quenching of energy-conserving flip-flops by the orientation-dependent quadrupole splitting is why spectral spin diffusion among deuterons is slow \cite{abragamgoldman1982,suter1985}, and why holes can be burnt and held in the deuteron line \cite{keller2017,keller2020}.

\paragraph*{Rotating sample.} Now $\Delta_{ij}(t)$ sweeps through zero whenever $c_i=c_j$ (same branch) or $c_i=-c_j$ (cross branch), which for a generic pair happens two to four times per turn. Each passage is a Landau--Zener crossing of the pair \cite{landau1932,zener1932}, with
\begin{equation}
P_{\rm swap}=1-\exp\Bigl(-\frac{2\pi V_{ij}^2}{\alpha_{ij}}\Bigr),\qquad
\alpha_{ij}=\Bigl|\frac{\dd\Delta_{ij}}{\dd t}\Bigr|\le2\alpha_{\rm rot},
\label{eq:LZpair}
\end{equation}
and an adiabatic crossing exchanges the two product states, i.e.\ swaps the complete single-transition states of the two spins. The crossing is adiabatic when $\Omega\ll\Omega^*$ and diabatic (no exchange) when $\Omega\gg\Omega^*$, where
\begin{equation}
\Omega^*=\frac{2\pi V_{ij}^2}{18\,\wQ\sin\beta}
\label{eq:Omegastar}
\end{equation}
is the rotation rate at which the exponent in Eq.~\eqref{eq:LZpair} is unity. For $\beta=54.74\degree$ this is $\Omega^*=7.4$~rad\,s$^{-1}$ (period $0.84$~s) for a geminal pair, $1.0$~rad\,s$^{-1}$ ($6$~s) at $2.5$~\AA{} and $0.35$~rad\,s$^{-1}$ ($18$~s) at $3$~\AA. At a $40$~s period the exponent for a geminal pair is $43$: every crossing is a complete swap, for all near neighbours, several times per turn. Table~\ref{tab:swap} gives the swap probability as a function of the rotation rate.

\begin{table*}
\caption{Landau--Zener swap probability for a geminal CD$_2$ pair ($V/2\pi=240$~Hz) at one crossing, Eq.~\eqref{eq:LZpair} with $\alpha_{ij}=\alpha_{\rm rot}$, $\beta=54.74\degree$; and the inversion fidelity of a fixed-frequency rf field of amplitude $\nu_1$ traversed by the same rotation, Eq.~\eqref{eq:LZ} with $\alpha=\alpha_{\rm rot}$.}
\label{tab:swap}
\begin{ruledtabular}
\begin{tabular}{lcccc}
$\Omega$ & equiv.\ period & $90\degree$ step & $P_{\rm swap}$ & rf fidelity ($\nu_1=5$~kHz) \\
\hline
$0.16$~rad/s & $40$~s & $10$~s & $1.00$ & $1.00$ \\
$6.3$ & $1$~s & $0.25$~s & $0.66$ & $1.00$ \\
$16$ & $0.4$~s & $0.1$~s & $0.38$ & $1.00$ \\
$52$ & $0.12$~s & $30$~ms & $0.13$ & $1.00$ \\
$157$ & $40$~ms & $10$~ms & $0.05$ & $1.00$ \\
$630$ & $10$~ms & $2.5$~ms & $0.01$ & $0.92$ \\
\end{tabular}
\end{ruledtabular}
\end{table*}

\subsection{What scrambling does to an inversion pattern}
\label{sec:scramble}

\paragraph*{Same-branch swaps.} A same-branch swap exchanges the $(+1,0)$ populations of two bonds and leaves the ensemble totals $n_+,n_0,n_-$, hence the ensemble $\Pzz$, unchanged. Its effect is on the \emph{pattern}: after many swaps the inversion is distributed at random over bonds irrespective of their orientation. The damage is sequential. Let $F_n$ be the inverted fraction after $n$ stops and let the pattern be fully randomized before stop $n+1$. The cap of stop $n+1$, of fractional area $f=1-\cos\theta_1$, then contains inverted bonds with probability $F_n$, and sweeping it un-inverts those while inverting the rest:
\begin{equation}
F_{n+1}=F_n+f\,(1-2F_n),\qquad F_N=\tfrac12\bigl[1-(1-2f)^N\bigr] .
\label{eq:recursion}
\end{equation}
The fixed point is $F=\tfrac12$, at which $\Pzz=\tfrac12(\Pzz^{\rm inv}+\Pzz^{\rm eq})$, exactly the saturation value of Table~\ref{tab:bound}. Scrambling therefore converts a sequence of inversions into saturation, as it must: a random permutation of the two populations of a transition over the ensemble is indistinguishable from their equalization. If only a fraction $s$ of the pattern is randomized between stops (the rest remaining in previously swept caps, which are disjoint from the next one), the recursion is $F_{n+1}=F_n+f(1-2sF_n)$, which interpolates between $F_N=Nf$ at $s=0$ and Eq.~\eqref{eq:recursion} at $s=1$. Table~\ref{tab:scramble} gives the result.

\begin{table*}
\caption{Inverted fraction $F$ and ensemble $\Pzz$ (route A from $P=0.5$) after the full sequence, for a fraction $s$ of the inversion pattern randomized by same-branch swaps between stops. $s=1$ is the regime of continuous rotation at the rates of Ref.~\cite{keller2020}; $s\simeq0.3$ corresponds to $90\degree$ transits of order $30$~ms (Table~\ref{tab:swap}, a few crossings per transit).}
\label{tab:scramble}
\begin{ruledtabular}
\begin{tabular}{lcccccc}
 & \multicolumn{2}{c}{$s=0$} & \multicolumn{2}{c}{$s=0.3$} & \multicolumn{2}{c}{$s=1$} \\
Scheme & $F$ & $\Pzz$ & $F$ & $\Pzz$ & $F$ & $\Pzz$ \\
\hline
four stops, $f=0.183$ & $0.734$ & $-0.571$ & $0.621$ & $-0.453$ & $0.420$ & $-0.242$ \\
six stops, $f=0.149$ & $0.896$ & $-0.740$ & $0.718$ & $-0.554$ & $0.441$ & $-0.264$ \\
\end{tabular}
\end{ruledtabular}
\end{table*}

\paragraph*{Cross-branch swaps.} These do change $\Pzz$. An adiabatic crossing exchanges the populations of $|0,0\rangle$ and $|{+1},{-1}\rangle$ (and, at the same crossing, of $|0,0\rangle$ and $|{-1},{+1}\rangle$, since both are degenerate with $|0,0\rangle$ when $c_i=-c_j$). For an uncorrelated pair the change in the pair's $m=0$ occupation is $\Delta n_0=-2[p_0(i)p_0(j)-p_+(i)p_-(j)]$, so per spin
\begin{equation}
\delta\Pzz=3\bigl[p_0(i)\,p_0(j)-p_+(i)\,p_-(j)\bigr]
\label{eq:crossloss}
\end{equation}
for the $|{+1},{-1}\rangle$ channel and the same with $p_+(i)p_-(j)\to p_-(i)p_+(j)$ for the other. On the single-temperature manifold, Eq.~\eqref{eq:detbal}, both brackets vanish: the DNP state is the fixed point, as found in Sec.~\ref{sec:relaxation} by rate equations. For an inverted bond meeting an uninverted neighbour the two channels give $\delta\Pzz=+0.40$ and $+0.28$ per spin, and for two inverted bonds $+1.05$, from a starting value of $-0.85$. One or two cross-branch crossings therefore undo the inversion of a bond entirely. Since the strongest-coupled partner of any inverted bond, its geminal neighbour at $109.5\degree$, is never inside the same cap, every inverted bond has at least one such partner, and at the rotation rates of Table~\ref{tab:swap} it meets it adiabatically within a fraction of a turn. Continuous slow rotation after an inversion thus relaxes the enhanced state to the spin-temperature curve within one turn, independently of the same-branch scrambling.

Both effects are absent in the static sample, where the crossing condition is never met except for the quenched set of pairs described above, and both are suppressed in fast rotation, where every crossing is diabatic.

\subsection{The timing window}
\label{sec:window}

The sequence therefore has two kinds of interval with incompatible speed requirements.

\paragraph*{Dwell.} During a sweep of length $T_p$ the sample must be at rest to the extent that the cap boundary moves by less than the band-edge width $\Omega_1/3\wQ$ in reduced units and the rotational chirp is negligible against the rf chirp:
\begin{equation}
\Omega\,T_p\,\sin\beta\ll\frac{\sqrt2\,\wone}{18\,\wQ},\qquad \alpha_{\rm rot}\ll\frac{\wone^2}{\mathcal A} .
\label{eq:dwell}
\end{equation}
For $T_p\simeq1$~ms and $\nu_1=5$~kHz both are satisfied for any $\Omega\lesssim30$~rad\,s$^{-1}$, and trivially at rest.

\paragraph*{Transit.} Between sweeps the sample must move fast enough that the pair crossings of Eq.~\eqref{eq:LZpair} are diabatic for the strongest couplings present,
\begin{equation}
\Omega\gg\Omega^*_{\rm geminal}\simeq7~\mathrm{rad\,s^{-1}} .
\label{eq:transit}
\end{equation}
A $90\degree$ step in $30$~ms ($\Omega=52$~rad\,s$^{-1}$) gives $P_{\rm swap}=0.13$ per crossing and, with a few crossings per transit, $s\simeq0.3$ in Table~\ref{tab:scramble}; a step in $10$~ms gives $0.05$ per crossing. Intermolecular pairs at $2.5$--$3$~\AA{} are diabatic already at $\Omega\gtrsim1$~rad\,s$^{-1}$. Stepping a target cell of several grams by $90\degree$ in tens of milliseconds in superfluid helium is a mechanical requirement outside the scope of this paper, but it is a requirement on a drive, not on the spin physics.

\paragraph*{Hold.} After the last sweep the sample stays at rest for the measurement and the beam. The quenched static rates of Sec.~\ref{sec:relaxation} then apply, and any further rotation re-opens both channels.

\paragraph*{Isotopic dilution.} Since $V_{ij}\propto r_{ij}^{-3}$, $\Omega^*\propto r^{-6}$. In a matrix deuterated to a fraction $x$ the mean D--D spacing scales as $x^{-1/3}$ and $\Omega^*\propto x^{2}$; geminal D--D pairs occur with probability $x^2$ per CH$_2$ group. At $x=0.1$ the transit condition relaxes by two orders of magnitude, to periods of order $10$--$100$~s, which the existing drive of Ref.~\cite{keller2020} already provides. Protons in the matrix do not transport deuteron order, since H--D flip-flops are forbidden by the Larmor mismatch; they contribute only to the local field and so shorten $T_2$, which enters the dwell condition through $\tau_p$ of Sec.~\ref{sec:inversion}. The cost is the deuteron density of the target, a figure-of-merit trade that the scattering experiment must weigh. The suppression of spin diffusion by dilution is the standard one of the spin-diffusion literature, where the diffusion constant scales with the coupling and the square of the spacing \cite{bloembergen1949,lowegade1967}.

\subsection{Simultaneous rf and rotation: a parity theorem}
\label{sec:parity}

\begin{figure}
\includegraphics[width=\columnwidth]{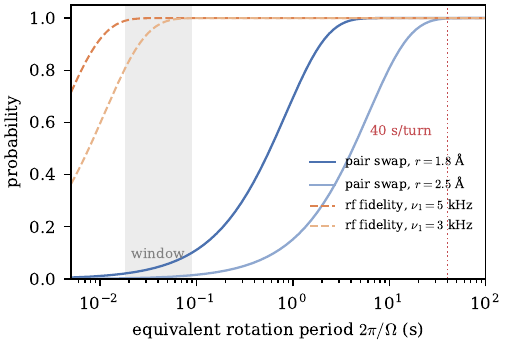}
\caption{Rotation-rate window for the transits, $\beta=54.74\degree$. Solid: Landau--Zener swap probability per pair crossing, Eq.~\eqref{eq:LZpair}, for geminal ($1.8$~\AA) and intermolecular ($2.5$~\AA) deuteron pairs; the transits must lie to the left, where crossings are diabatic. Dashed: inversion fidelity of a fixed-frequency rf field traversed by the rotation, Eq.~\eqref{eq:LZ}, for $\nu_1=3$ and $5$~kHz. The shaded band is the window of Eq.~\eqref{eq:windowchain}, in which rf passage would be adiabatic while pair crossings are diabatic; by the parity theorem of Sec.~\ref{sec:parity} nothing is gained by using it. The marker at $40$~s is the rotation of Ref.~\cite{keller2020}.}
\label{fig:window}
\end{figure}

One might hope to combine the two intervals, letting the rotation itself sweep each bond through a fixed-frequency rf field while the motion is fast enough to be diabatic for pairs. The two requirements are compatible: a fixed-frequency rf field traversed by the rotation inverts a transition with fidelity $1-\exp(-\pi\wone^2/\alpha_{\rm rot})$, Eq.~\eqref{eq:LZ}, so the condition for the rf passage to be adiabatic while the pair crossings are diabatic is
\begin{equation}
V_{ij}^2\ll\alpha_{\rm rot}\ll\wone^2 ,
\label{eq:windowchain}
\end{equation}
a window of width $(\wone/V)^2\simeq400$ for $\nu_1=5$~kHz and geminal pairs, corresponding to equivalent periods between about $20$ and $90$~ms (Fig.~\ref{fig:window}). The passage adiabaticity $\wone^2/\alpha_{\rm rot}=\wone^2/18\wQ\Omega\sin\beta$ has the form of Vega's parameter $\wone^2/\wQ\omega_r$ for quadrupolar spins under sample spinning \cite{vega1992}, and the process is the slow-spinning analogue of rotor-assisted population transfer \cite{yao2000}. The window exists; it does not help, for the following reason.

\paragraph*{Theorem.} Let the rf frequency $\omega_{\rm rf}(t)$ and the transition frequency $\omega_i(t)$ of bond $i$ be continuous on $[t_a,t_b]$, let $\Delta_i(t)=\omega_{\rm rf}(t)-\omega_i(t)$ be nonzero at the endpoints, and assume (A1) every zero of $\Delta_i$ is simple and traversed adiabatically, (A2) no zero of $\Delta_i$ coincides with a pair crossing of bond $i$ or with a zero of $\Delta$ for its other transition, and (A3) away from its zeros $|\Delta_i|\gg\Omega_1$, so that the populations of the transition are constant there. Then at $t_b$ the two populations of the transition are exchanged if $\operatorname{sgn}\Delta_i(t_a)\ne\operatorname{sgn}\Delta_i(t_b)$ and unchanged otherwise. The set of bonds net-inverted by the episode is
\begin{equation}
S=\bigl\{\,i:\ \operatorname{sgn}\Delta_i(t_a)\ne\operatorname{sgn}\Delta_i(t_b)\,\bigr\} ,
\label{eq:parityset}
\end{equation}
and depends on the history only through its endpoints.

\paragraph*{Proof.} By (A1)--(A3) the populations change only at zeros of $\Delta_i$, and at each zero they are exchanged. The number of simple zeros of a continuous function on an interval with nonzero endpoint values is even if the endpoint values have the same sign and odd otherwise (intermediate value theorem applied between consecutive zeros). An even number of exchanges is the identity. $\square$

\paragraph*{Corollaries.} Write $C(\hat n,\theta)$ for the pair of antipodal caps of half-angle $\theta$ about the field direction $\hat n$, and $\theta(R)$ for the cap angle at which a transition sits at reduced frequency $R$, Eq.~\eqref{eq:theta1}.
\begin{enumerate}
\item \emph{Static sweep.} With the sample at rest and the rf swept from $R_a=-2$ to $R_b=-R_1$ on the $(+1,0)$ branch, $S=C(\hat n,\theta(R_1))$: the cap of Sec.~\ref{sec:rotation}. The theorem reproduces the selective inversion of Sec.~\ref{sec:inversion}.
\item \emph{Fixed rf, rotation from $\hat n_a$ to $\hat n_b$.} With $\omega_{\rm rf}$ fixed at reduced frequency $R_{\rm rf}$ in the clean band, $S=C(\hat n_a,\theta_{\rm rf})\,\triangle\,C(\hat n_b,\theta_{\rm rf})$, the symmetric difference of the initial and final caps with $\theta_{\rm rf}=\theta(R_{\rm rf})$. Every bond that the moving cap passes over is crossed twice.
\item \emph{Concatenation.} A chain of such episodes through stops $\hat n_1,\dots,\hat n_N$ with the rf on throughout gives $\bigtriangleup_k\bigl[C_k\triangle C_{k+1}\bigr]=C_1\triangle C_N$: the intermediate caps cancel in pairs. A full turn gives $S=\varnothing$.
\item \emph{rf swept in during rotation.} If $\omega_{\rm rf}$ starts outside the branch ($R_a<-2$) and ends at $R_b$ while the sample moves from $\hat n_a$ to $\hat n_b$, then $S=C(\hat n_b,\theta(R_b))$: the same as a static sweep at the final orientation.
\end{enumerate}
A single continuous rf episode therefore inverts at most two caps, and $N$ caps require $N$ episodes separated by intervals in which the rf is off, which is the stepped sequence of Sec.~\ref{sec:rotation}. The hit-once condition of Eq.~\eqref{eq:F} is the discrete form of the theorem. Applying the rf during the transits is at best harmless, provided it is parked strictly inside the clean band (at $R=-1$ exactly it sits on the horn of the other branch, the most populous orientation in the sample) and provided every crossing is fully adiabatic; a partially adiabatic crossing, Eq.~\eqref{eq:LZ} with $\pi\wone^2/\alpha_{\rm rot}$ of order unity, leaves the bond in a mixture that no later pulse can repair, which is the practical reason to keep the rf off while the sample moves.

The theorem also explains the two results of Sec.~\ref{sec:rotation} for continuous rotation: a bond that enters the clean band leaves it, so continuous rotation with a continuously swept rf inverts nothing net, whereas continuous rotation with continuous \emph{saturation}, as in Ref.~\cite{keller2020}, is unaffected because saturation is idempotent.

\subsection{The sequence}
\label{sec:sequence}

With these constraints the four-stop, single-axis sequence (route A) is:
\begin{enumerate}
\item Mount the rotation axis at $\beta=54.74\degree$ to $B_0$ with a drive capable of $90\degree$ steps in $\lesssim30$~ms and an encoder good to $1\degree$.
\item DNP to $P\simeq0.5$ with the sample at rest; microwaves off.
\item Stop 1, sample at rest: isolate the Q-meter \cite{court1993}, fire one adiabatic sweep over $-2\le R\le-1$ (full pedestal of the $(+1,0)$ branch, $T_p\simeq1$~ms), restore the Q-meter.
\item Step $90\degree$ in $\lesssim30$~ms, rf off; settle.
\item Stops 2, 3, 4: repeat steps 3--4. Total elapsed time since step 3 is well under one second.
\item Do not rotate again. Measure $\Pzz$ from the calibrated area difference $C(I_+-I_-)$ with the fit allowed to go negative over the swept band; expected $-0.57$ for diabatic transits, $-0.45$ for $s\simeq0.3$ (Table~\ref{tab:scramble}).
\item Hold at rest with the beam on until $|\Pzz|$ has fallen to the chosen fraction of its peak; return to step 2.
\end{enumerate}
The six-stop variant uses $\beta=63.43\degree$, $72\degree$ steps, the band $-2\le R\le-1.17$, and a tilt of the axis onto $B_0$ for the sixth sweep; it adds one transit and a tilt mechanism for a gain from $-0.57$ to $-0.74$ and is best attempted after the four-stop sequence has been demonstrated. Route B is either sequence with the sweep on the $(0,-1)$ pedestal.

\section{Relaxation of the enhanced state}
\label{sec:relaxation}

\subsection{Flip-flop channels and their fixed point}

The flip-flop term of the secular dipolar coupling, Eq.~\eqref{eq:HD}, conserves total $I_z$ and connects $|m_i,m_j\rangle$ to $|m_i+1,m_j-1\rangle$. In a Zeeman-dominated solid the process is energy conserving only to within the dipolar linewidth, and the quadrupolar splitting of Eq.~\eqref{eq:transitions} makes energy conservation a condition on bond orientations \cite{abragam1961,abragamgoldman1982}. For spin-1 pairs there are two kinds of allowed process.
\begin{enumerate}
\item \emph{Same-branch} flip-flops, $0\to+1$ on $i$ with $+1\to0$ on $j$, allowed when $\omega_{+1,0}(i)\simeq\omega_{+1,0}(j)$. These transport population between bonds within one branch and conserve the ensemble totals $n_+,n_0,n_-$. This is spectral spin diffusion \cite{suter1985}; it smooths the inverted/uninverted pattern over orientations but leaves the ensemble $\Pzz$ unchanged.
\item \emph{Cross-branch} flip-flops, $0\to+1$ on $i$ with $0\to-1$ on $j$, allowed when $\omega_{+1,0}(i)\simeq\omega_{0,-1}(j)$, i.e.\ when the two bonds' different transitions coincide in frequency. The net effect is $n_0\to n_0-2$, $n_+\to n_++1$, $n_-\to n_-+1$: $P$ is conserved and $\Pzz$ changes by $-6/N$ per event.
\end{enumerate}
The cross-branch channel runs forward at a rate proportional to $p_0(i)p_0(j)$ and backward, both spins returning to $m=0$, at a rate proportional to $p_+(i)p_-(j)$. Its stationary condition for bonds of equal populations is
\begin{equation}
p_0^2=p_+p_-,
\end{equation}
Eq.~\eqref{eq:detbal}: the channel relaxes the alignment order at fixed Zeeman order onto the single-temperature manifold. This is the cross-relaxation of Bloembergen \emph{et al.} \cite{bloembergen1959} between the Zeeman and alignment reservoirs of Goldman's description \cite{goldman1970}, it is the microscopic content of the statement that an rf-burnt hole refills \cite{keller2020}, and it is the rate-equation form of the per-crossing result, Eq.~\eqref{eq:crossloss}, which vanishes on the same manifold. After route A the ensemble vector polarization is $\simeq0.19$, so the end point is $\Pzz^{\rm eq}\simeq+0.03$ and the enhanced state decays from $-0.74$ through zero toward that value.

\subsection{Rates}

The cross-branch coincidence in reduced units is $(3\cos^2\theta_i-1)=-(3\cos^2\theta_j-1)$, a condition met within the dipolar width $b/3\wQ\approx0.01$ of a measure-zero set of pair orientations. Nearest-neighbour deuterons on one molecule have fixed relative bond directions, so for most molecular orientations no near pair is matched and transfer proceeds through chains of weakly matched spins. This suppression by the ratio of the dipolar to the quadrupolar width is the same mechanism that makes deuteron spin diffusion slow \cite{abragamgoldman1982,suter1985}, and it is the precondition on which the hole-burning experiments of Refs.~\cite{keller2017,keller2020} rely. It makes $T_{1Q}$ long compared with $1/b\approx0.3$~ms, but how long is a property of the material and the paramagnetic centres at $1$~K that must be measured.

An upper bound follows from the paramagnetic relaxation that limits $T_{1Z}$. Relaxation by a randomly fluctuating magnetic field, the rank-1 perturbation through which the electron spins act on the nuclei \cite{abragamgoldman1978,wenckebach2016}, drives only $\Delta m=\pm1$ transitions. If both single-quantum transitions relax at the same rate $W$, which holds when the spectral density is flat over the $\sim100$~kHz separating them, the rate equations $\dot p_\pm=W(p_0-p_\pm)$, $\dot p_0=W(p_++p_--2p_0)$ give
\begin{equation}
\dot P=-W P,\qquad \dot\Pzz=-3W\,\Pzz,\qquad T_{1Q}=\tfrac13 T_{1Z},
\label{eq:T1Q}
\end{equation}
the well-known factor of three between Zeeman and alignment relaxation for single-quantum mechanisms \cite{spiess1980}. The cross-branch flip-flop channel adds to this, so $T_{1Q}\lesssim T_{1Z}/3$. Since the DNP build-up time of the material is tens of minutes \cite{keller2020} and $T_{1Z}$ with microwaves off at $1$~K is typically of the same order or longer, $T_{1Q}$ is plausibly in the range of ten minutes to an hour, which is exactly the range in which the break-even of Sec.~\ref{sec:cycle} falls. Two practical points follow. Rotation after the sequence sweeps pairs through the coincidence condition at the adiabatic rates of Sec.~\ref{sec:paircrossing} and must be stopped; the static rates discussed here apply only to a sample at rest. DNP microwaves drive both transitions to a common spin temperature at the DNP rate and so also restore Eq.~\eqref{eq:detbal}; whether they stay on during the hold depends on whether the loss of Zeeman order or of alignment order is the larger penalty, which the same measurement settles.

\subsection{Duty cycle and break-even}
\label{sec:cycle}

\begin{figure}
\includegraphics[width=\columnwidth]{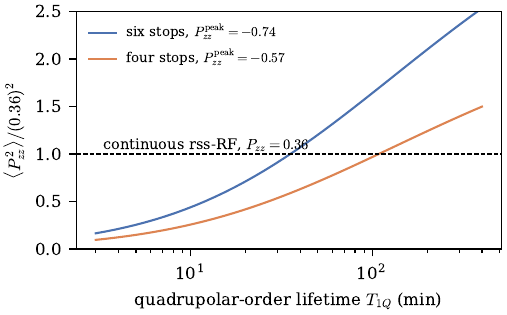}
\caption{Time-averaged figure of merit $\langle\Pzz^2\rangle$ of the pulsed scheme relative to continuous rotating semi-saturation ($\Pzz=0.36$, dashed), as a function of the alignment-order lifetime $T_{1Q}$, for $30$~min of DNP re-preparation per cycle and a hold time optimized at each $T_{1Q}$.}
\label{fig:fom}
\end{figure}

Repolarizing from the post-inversion $P\simeq0.19$ back to $0.50$ takes about $2.8$ DNP time constants, roughly $30$~min for the material of Ref.~\cite{keller2020}. With $\Pzz(t)=\Pzz^{\rm eq}+(\Pzz^{\rm A}-\Pzz^{\rm eq})e^{-t/T_{1Q}}$ during a hold of length $T_h$ optimized for each $T_{1Q}$, the time-averaged figure of merit $\langle\Pzz^2\rangle$ over the cycle compares with the continuous semi-saturation value $0.36^2=0.130$ as shown in Fig.~\ref{fig:fom}. The six-stop scheme breaks even at $T_{1Q}\simeq35$~min and the four-stop scheme at $\simeq100$~min. A shorter re-preparation, or a partial cycle that re-inverts before full repolarization, the bath-contact step of Sec.~\ref{sec:bound}, lowers both thresholds; the measurement of $T_{1Q}$ decides whether the pulsed scheme is worth building.

\section{Why magic-angle spinning and recoupling do not help}
\label{sec:mas}

It is natural to ask whether fast sample spinning, which removes the orientation dependence of Eq.~\eqref{eq:HQ1} \cite{andrew1958,lowe1959,maricq1979}, combined with rotor-synchronized recoupling sequences \cite{gullion1989,hing1992}, could replace the orientation indexing of Sec.~\ref{sec:rotation}. Average Hamiltonian theory \cite{haeberlen1968,mehring1983} answers this in the negative on two counts.

Under rotation about an axis at angle $\beta$ to $B_0$ the spatial part of $H_Q^{(1)}$, a traceless rank-2 tensor, becomes time dependent,
\begin{equation}
\begin{aligned}
H_Q^{(1)}(t)&=\wQ\Bigl[A_0+\sum_{k=\pm1,\pm2}A_k\,e^{ik\Omega t}\Bigr](3I_z^2-2),\\
A_0&\propto P_2(\cos\beta).
\end{aligned}
\end{equation}
with coefficients $A_k$ depending on the bond orientation relative to the rotor \cite{maricq1979}. The zeroth-order average Hamiltonian over a rotor period retains only $A_0$, which vanishes at the magic angle. The two transitions then become degenerate in the averaged picture, $\bar H=-\wD I_z$, and the only operations an rf field can generate are global rotations $\exp(-i\phi\,\hat n\cdot\bm I)$ in the spin-1 representation of SU(2). Such rotations map the population vector by the Wigner matrices $d^{(1)}$, whose only population permutation is $p_+\leftrightarrow p_-$; they cannot move population into or out of $m=0$ in the way Table~\ref{tab:bound} requires. The quadrupolar splitting is thus not an obstacle to be averaged away but the resource that makes the two transitions distinguishable.

Recoupling restores a scaled copy of the averaged interaction by rotor-synchronized pulses. But the spatial tensor of $H_Q^{(1)}$ transforms under SO(3) in the $\ell=2$ representation, and any first-order average over spatial rotations and rf rotations is a linear map of its components; it cannot generate an $\ell=0$ (isotropic) part from an $\ell=2$ tensor whose isotropic part is zero by the Laplace condition $\sum_\alpha V_{\alpha\alpha}=0$. Every recoupled first-order quadrupolar Hamiltonian is therefore still a powder pattern, scaled or component-selected, and the overlap of the two branches that the present scheme circumvents by geometry reappears in it. Second-order terms do contain an isotropic quadrupolar shift, but of order $\wQ^2/\wD\approx1$~kHz, comparable to the dipolar width and far below the $\sim60$~kHz that branch selectivity requires. REDOR and TEDOR \cite{gullion1989,hing1992} recouple heteronuclear dipolar couplings for distance measurement, a different interaction altogether.

The engineering reinforces the physics. Magic-angle spinning at the kHz rates needed to average a $250$~kHz-wide pattern requires rotors of millimetre scale on gas bearings; DNP-MAS has been pushed to $\simeq25$~K in such rotors \cite{thurber2008,maly2008}, not to $1$~K in a $15$~cm$^3$ cell immersed in superfluid helium with a microwave horn and a particle beam. At the sub-hertz rates available to a target the rotor period exceeds every coherence time of the spin system by orders of magnitude, so rotor-synchronized coherent sequences have nothing to act on. The analogues that do carry over are the population-transfer methods developed for half-integer quadrupolar nuclei under slow spinning, double frequency sweeps \cite{iuga2000}, rotor-assisted population transfer \cite{yao2000} and hyperbolic-secant inversion of satellite transitions \cite{siegel2004}, all of which are selective adiabatic inversions of the kind used in Sec.~\ref{sec:inversion}.

\section{Results and proposed measurements}
\label{sec:tests}

\begin{table*}
\caption{Ensemble tensor polarization after one pass from $P=0.5$, $\Pzz=F\,\Pzz^{\rm inv}+(1-F)\times0.197$, compared with the measured values of Ref.~\cite{keller2020} and the ceilings of Table~\ref{tab:bound}. The first three rows assume diabatic transits (Sec.~\ref{sec:window}); the two rows marked $s=1$ are the same sequences with slow transits that fully scramble the inversion pattern, Eq.~\eqref{eq:recursion}, and ignore the additional cross-branch loss.}
\label{tab:results}
\begin{ruledtabular}
\begin{tabular}{lcccc}
Scheme & $F$ & $\Pzz$, route A & $\Pzz$, route B & $P$ after A \\
\hline
six icosahedral stops & $0.896$ & $-0.740$ & $+0.604$ & $0.19$ \\
five icosahedral stops & $0.747$ & $-0.584$ & $+0.537$ & $0.24$ \\
four cube-diagonal stops & $0.734$ & $-0.571$ & $+0.531$ & $0.24$ \\
six stops, transits at $40$~s/turn ($s=1$) & $0.441$ & $-0.264$ & $+0.332$ & -- \\
four stops, transits at $40$~s/turn ($s=1$) & $0.420$ & $-0.242$ & $+0.317$ & -- \\
rotating semi-saturation \cite{keller2020} & -- & -- & $+0.36$ & $0.43$ \\
semi-saturation, static \cite{keller2020} & -- & -- & $+0.27$--$0.31$ & $0.50$ \\
saturation ceiling & $1$ & $-0.326$ & $+0.424$ & -- \\
unitary bound, Eq.~\eqref{eq:bound} & $1$ & $-0.849$ & $+0.652$ & -- \\
\end{tabular}
\end{ruledtabular}
\end{table*}

Table~\ref{tab:results} summarizes the single-pass results. The contrast between the diabatic-transit rows and the $s=1$ rows is the central practical point of Sec.~\ref{sec:timing}: with the slow continuous rotation used for semi-saturation, the pulsed sequence would deliver no more than saturation does, and with the cross-branch loss of Eq.~\eqref{eq:crossloss} it would deliver less. The inverted values carry the Landau--Zener fidelity of Eq.~\eqref{eq:LZ} and lose $\approx1.5\%$ of each cap to the band edge, so a realistic six-stop figure is $-0.72$; with the $0.80$ efficiency measured for full-line AFP in Ref.~\cite{hossain2026} as a pessimistic bound the figure would be $-0.56$, still well beyond the saturation ceiling.

The inverted bonds emit rather than absorb on the swept transition, so one branch of the CW-NMR line goes partly negative. The populations are far from the single-temperature manifold, so the intensity-ratio formulas do not apply; $\Pzz$ must be taken from the calibrated area difference $C(I_+-I_-)$, valid for any population distribution \cite{dulya1997,keller2013}, with a fit that allows the two intensities to carry opposite signs over the swept band. The thermal-equilibrium calibration of $C$ is unaffected.

Four measurements decide the design, in order of leverage.
\begin{enumerate}
\item $T_{1Q}$ at $1$~K and $5$~T, microwaves off and on, from the decay of alignment order prepared either by a Jeener--Broekaert pair \cite{jeener1967,spiess1980} or by a single clean-band sweep on a static sample, read out by the area difference versus delay. This sets the hold time and the break-even of Fig.~\ref{fig:fom}, and tests the bound $T_{1Q}\le T_{1Z}/3$ of Eq.~\eqref{eq:T1Q}.
\item A single-stop test on a static sample. One cap covers $1-\cos\theta_c=18\%$ of the bonds; route A should give $\Pzz=0.18\times(-0.849)+0.82\times0.197=+0.01$, a clean null, and route B $+0.28$. This checks the inversion, the band edge and the fit before any rotation hardware is built, and a residual signal localizes any component of the line with a different $\wQ$ (a motionally narrowed CD$_3$ site or an O--D site in a C$_4$D$_9$OD sample), whose clean region lies at a different absolute frequency.
\item The rotation-driven swap rate, by inverting one cap at rest, rotating by $90\degree$ at a set speed, rotating back, and reading the surviving area difference as a function of the transit time. This maps $P_{\rm swap}$ of Table~\ref{tab:swap} directly, including the cross-branch loss, and sets the drive specification; the same measurement at two deuteration levels tests the $r^{-6}$ scaling of Sec.~\ref{sec:window}.
\item The deuteron $T_2$ from a Hahn echo \cite{hahn1950} on the pedestal, to confirm $\tau_p\approx0.2$~ms is short enough or to set the required $\wone$.
\item $\nu_1$ in the coil at $20$--$50$~W from the nutation rate of a pedestal-selective pulse, which fixes $T_p$ through Eq.~\eqref{eq:Tp}.
\end{enumerate}

\section{Conclusion}
\label{sec:conclusion}

Starting from the first-order quadrupole Hamiltonian, the spin-temperature description of Zeeman and alignment order, and the single-transition operator formalism, we have derived the largest tensor polarization that rf manipulation and sample rotation can extract from a dynamically polarized spin-1 target, the sequence that attains it, and the mechanism that limits its lifetime. The bound is a majorization statement: from $P=0.5$ no unital operation can take $\Pzz$ outside $[-0.85,+0.65]$, saturation reaches only half the accessible transfer, and selective $\pi$ rotations reach the bound. In a powder the selectivity is available only in the pedestal region of the Pake doublet, which turns the problem into a packing of caps on the orientation sphere whose solutions are the cube diagonals, with the rotation axis at the magic angle, and the icosahedral axes; adiabatic inversion of the clean band at each stop then yields $\Pzz=-0.57$ (four stops, one axis) or $-0.74$ (six stops, two axes) in a single pass, against $0.36$ by semi-saturation. Rotation is not free: it chirps every transition and turns the dipolar flip-flops that the quadrupole splitting quenches at rest into adiabatic pair crossings, which scramble an inversion pattern into saturation and relax the alignment order within a turn at the slow rates used for semi-saturation. The transits between stops must therefore be fast enough to be diabatic for the strongest deuteron pairs, which isotopic dilution relaxes as $r^{-6}$, and a parity theorem shows that leaving the rf on during the motion cannot recover anything, because the net effect of any continuous episode of adiabatic passage is fixed by its endpoints. At rest the enhanced state decays through the quenched cross-branch channel, whose fixed point is the spin-temperature relation $p_0^2=p_+p_-$, with a lifetime bounded by $T_{1Z}/3$; whether the scheme improves the figure of merit of a scattering experiment therefore depends on one measurable number, $T_{1Q}$, for which we have given the break-even. Average Hamiltonian theory shows that magic-angle spinning and recoupling cannot substitute for this geometry, because the first-order quadrupole coupling is a traceless rank-2 spatial tensor that no averaging renders orientation independent while preserving the branch selectivity on which population transfer into or out of $m=0$ depends.

\appendix

\section{Pake densities and the branch overlap}
\label{app:pake}

With $x=\cos\theta$ uniform on $[0,1]$ and $R_{0,-1}=3x^2-1$, the density per unit $R$ is $g_-(R)=\dd x/\dd R=1/(6x)=1/(2\sqrt{3(1+R)})$ on $[-1,2]$, and by the mirror symmetry $g_+(R)=g_-(-R)$. Both integrate to unity. The overlap of the two branches,
\begin{equation}
\int_{-1}^{1}g_+(R)\,g_-(R)\,\dd R=\frac{1}{12}\int_{-1}^{1}\frac{\dd R}{\sqrt{1-R^2}}=\frac{\pi}{12}\approx0.26,
\end{equation}
is the spectral factor entering the cross-branch channel of Sec.~\ref{sec:relaxation} before the orientation-matching suppression; the same-branch self-overlap $\int g_\pm^2\,\dd R$ diverges logarithmically at the horn and is cut off by the dipolar width.

\section{Separation of stops on a cone}
\label{app:cone}

For stops at azimuths $\varphi_k=2\pi k/N$ on the cone \eqref{eq:cone}, $\hat n_j\cdot\hat n_k=\cos^2\beta+\sin^2\beta\cos(\varphi_j-\varphi_k)$, which gives Eq.~\eqref{eq:sadj} for adjacent stops and $\cos s_{\rm 2nd}=\cos^2\beta+\sin^2\beta\cos(4\pi/N)$ for second neighbours. The pole satisfies $\hat a\cdot\hat n_k=\cos\beta$. For $N=5$ the two constraints $s_{\rm adj}\ge2\theta_1$ and $\pi-s_{\rm 2nd}\ge2\theta_1$ are both saturated at $\cos^2\beta=\tfrac15$; a one-dimensional search in $\beta$ confirms that this is the maximum of $\min(s_{\rm adj},\pi-s_{\rm 2nd})$ and that adding the pole does not reduce it.

\section{Monte Carlo check}
\label{app:mc}

The coverage fractions of Table~\ref{tab:caps} were verified by drawing $3\times10^5$ unit vectors from the isotropic distribution, counting for each the number of stop lines $\hat n_k$ with $|\hat u\cdot\hat n_k|\ge\cos\theta_1$, and recording the fractions hit exactly once and more than once. The results agree with the closed forms $N(1-\cos\theta_1)$ to three figures, and show zero double hits for the cube-diagonal and icosahedral configurations and $37\%$ double hits for four stops at $45\degree$ on a perpendicular axis.

\section{Pair crossing Hamiltonians}
\label{app:pair}

For two spin-1 nuclei with first-order quadrupole energies $E_m^{(k)}=-\wD m+\wQ c_k(3m^2-2)$, $k=i,j$, the product states connected by the flip-flop operator $I_i^+I_j^-+I_i^-I_j^+$ and their energy differences are, with $\langle{+1}|I^+|0\rangle=\langle0|I^+|{-1}\rangle=\sqrt2$,
\begin{align}
&|{+1},0\rangle\leftrightarrow|0,{+1}\rangle:&&\langle0,{+1}|I_i^-I_j^+|{+1},0\rangle=2,\nonumber\\
&&&E_{+1,0}-E_{0,+1}=-3\wQ(c_i-c_j);\nonumber\\
&|0,{-1}\rangle\leftrightarrow|{-1},0\rangle:&&\langle{-1},0|I_i^-I_j^+|0,{-1}\rangle=2,\nonumber\\
&&&E_{0,-1}-E_{-1,0}=+3\wQ(c_i-c_j);\nonumber\\
&|0,0\rangle\leftrightarrow|{+1},{-1}\rangle:&&\langle{+1},{-1}|I_i^+I_j^-|0,0\rangle=2,\nonumber\\
&&&E_{+1,-1}-E_{0,0}=3\wQ(c_i+c_j);\nonumber\\
&|0,0\rangle\leftrightarrow|{-1},{+1}\rangle:&&\langle{-1},{+1}|I_i^-I_j^+|0,0\rangle=2,\nonumber\\
&&&E_{-1,+1}-E_{0,0}=3\wQ(c_i+c_j).
\end{align}
The first two are the same-branch channels on the upper and lower branch respectively; the last two are the cross-branch channels, degenerate with each other, so that at $c_i=-c_j$ the three states $|0,0\rangle$, $|{+1},{-1}\rangle$, $|{-1},{+1}\rangle$ form a three-level crossing in which $|0,0\rangle$ couples to the symmetric combination with matrix element $-\tfrac{1}{4}b_{ij}\cdot2\sqrt2$. With the flip-flop prefactor $-\tfrac14b_{ij}$ of Eq.~\eqref{eq:HD} each two-level coupling is $V_{ij}=\tfrac12|b_{ij}|$, which gives Eq.~\eqref{eq:Hpair}. Two-quantum pair processes such as $|{+1},{-1}\rangle\leftrightarrow|{-1},{+1}\rangle$ require the non-secular terms of the dipolar coupling and are suppressed by $(b/\wD)^2$. The $I_{zi}I_{zj}$ term of Eq.~\eqref{eq:HD} shifts all product states by amounts of order $b$ and contributes to the local width that sets the static matching condition of Sec.~\ref{sec:paircrossing} but does not drive transitions.

Under rotation, $c_k(t)$ follows Eq.~\eqref{eq:costheta_t} for each bond with its own $(\alpha_u,\phi_u)$, and $\Delta_{ij}(t)$ in Eq.~\eqref{eq:Deltas} is a smooth periodic function of $\Omega t$ whose zeros are the crossings. For a geminal pair, $\hat u_i\cdot\hat u_j=\cos109.5\degree$, the two trajectories $\cos\theta_i(t)$ and $\cos\theta_j(t)$ have the same period and different phases and amplitudes, and $c_i-c_j$ and $c_i+c_j$ each have at least two zeros per turn for generic orientation of the pair relative to the rotor. The crossing rate $\alpha_{ij}=|\dot\Delta_{ij}|=3\wQ|\dot c_i\mp\dot c_j|$ is bounded by $2\alpha_{\rm rot}$ and is of order $\alpha_{\rm rot}$ for generic crossings; the entries of Table~\ref{tab:swap} use $\alpha_{\rm rot}$.

\bibliographystyle{apsrev4-2}
\bibliography{refs}

\end{document}